\documentclass[twocolumn,showpacs,preprintnumbers,amsmath,amssymb,aps,floatfix,prl,superscriptaddress]{revtex4}
\usepackage{amsfonts}
\usepackage{dcolumn}
\usepackage{bm}
\usepackage{color}
\usepackage{epsfig}
\usepackage{hyperref}
\usepackage{bbold}

\begin{document}

\def\makered#1{{\color{red} #1}}
\def\makeblue#1{{\color{blue} #1}}

\newcommand{\be}{\begin{equation}}
\newcommand{\ee}{\end{equation}}
\newcommand{\bea}{\begin{eqnarray}}
\newcommand{\eea}{\end{eqnarray}}
\newcommand{\br}{\mathbf{r}}

\title{Local Control and $v$-Representability of Correlated Quantum Dynamics}

\author{S.E.B.~Nielsen}
\affiliation{Lundbeck Center for Theoretical Chemistry, Department of Chemistry, Aarhus University, 8000 Aarhus C, Denmark}
\author{M.~Ruggenthaler}
\affiliation{Department of Physics, Nanoscience Center, University of Jyv\"askyl\"a, 40014 Jyv\"askyl\"a, Finland}
\author{R.~van Leeuwen}
\affiliation{Department of Physics, Nanoscience Center, University of Jyv\"askyl\"a, 40014 Jyv\"askyl\"a, Finland}
\affiliation{European Theoretical Spectroscopy Facility (ETSF)}

\date{\today}

\begin{abstract}
We present a local control scheme to construct the external potential $v$ that, for a given initial state,
produces a prescribed time-dependent density in an interacting quantum many-body system.
The numerical method is efficient and stable even for large and rapid density variations
irrespective of the initial state and the interactions.
The method can at the same time be used to answer fundamental $v$-representability questions in density-functional theory.
In particular, in the absence of interactions,
it allows us to construct the exact time-dependent Kohn-Sham potential for arbitrary initial states.
We illustrate the method in a correlated one-dimensional two-electron system
with different interactions, initial states and densities.
For a Kohn-Sham system with a correlated initial state we demonstrate the interplay
between memory and initial-state dependence as well as the failure of any adiabatic approximation.
\end{abstract}
\maketitle

The $v$-representability question is one of the outstanding problems in density functional theory (DFT)
\cite{DFT, Chayes1985, Lammert2010}.
In time-dependent DFT (TDDFT) \cite{RvL1999, Baer2008, Li2008, Verdozzi2008, KurthS2011,CarstenBook, GFPP, Farzanehpour2012},
the question is whether, for a given initial state, there exists a local external potential $v$
that yields a prescribed density by solution of the time-dependent Schr\"odinger equation (TDSE).
In the case of a non-interacting system, $v$-representability amounts
to the existence of a Kohn-Sham (KS) system \cite{DFT, CarstenBook}.
The KS system has played a major role in the study of correlated many-body systems
as it allows for the treatment of interacting systems in an effective one-particle framework.
This feature greatly reduces computational costs \cite{Andrade2012}
and (TD)DFT has hence been one of the leading methods in electronic structure theory \cite{Burke2012}.
In practice the accuracy of the method is limited by the approximate nature of the density functionals that are used.
In TDDFT the most commonly used density functionals are based on the adiabatic approximation
in which the KS potential only depends on the instantaneous density.
These functionals can, however, fail in important cases \cite{CarstenBook} and therefore there is
a great need for better functionals.
To develop and benchmark such new functionals the availability of exact time-dependent KS potentials is highly desirable.
Although such potentials can be constructed in special cases
 \cite{Lein2005,Godby2012} no general practical scheme has been available so far.
In this Letter we provide such a scheme based on a recently introduced fixed-point formulation of TDDFT \cite{GFPP, GFPP 1-D}.
It is at the same time an efficient local control scheme based on the density which
augments other important control methods \cite{OCT1, OCT2, LCT1,LCT2} already of extensive use
in laser physics,
quantum optics \cite{Mancini2005} and the physics of ultracold gases \cite{CHU02}.
The scheme is closely related to existing methods \cite{LCT1,LCT2} but targets a spatially extended quantity instead.
It is applicable to general interactions and initial states and can deal with fast and large density changes.
We demonstrate the approach for an interacting one-dimensional
two-electron system with different interactions, initial states and densities.
For a KS system with a non-separable initial state we illustrate the connection
between memory and initial state dependence as well as the failure of any adiabatic approximation.

{\em The global fixed point method.}
We consider a $N$-electron system with a time-dependent Hamiltonian $\hat{H}(t) = \hat{T} + \hat{V}(t)+ \hat{W}$,
where $\hat{T}$ is the kinetic energy, $\hat{V} (t)$ the time-dependent external potential
and $\hat{W}$ the many-body interaction (which may even be time-dependent).
The expectation values $n (\br t)$ and $\mathbf{j}(\br t)$ of the density and current operators
(atomic units are used throughout)
\bea
\hat{n} (\br) &=& \sum_{l=1}^N \delta (\br -\br_l)  \nonumber \\
\hat{\mathbf{j}} (\br) &=&
\frac{1}{2 i} \sum_{l=1}^{N} \left(  \delta (\br -\br_l) \overrightarrow{\nabla}_{l} - 
 \overleftarrow{\nabla}_{l} \delta (\br -\br_l)  \right) \nonumber
\eea
satisfy equations of motion given by
\bea				
\partial_t n(\br t) &=& - \overrightarrow{\nabla} \cdot \mathbf{j} (\br t) 
\label{cont1}\\
\partial_t \mathbf{j} (\br t) &=& - n(\br t) \overrightarrow{\nabla} v (\br t) + \mathbf{Q} (\br t).
\label{cont2}
\eea
Here the internal local force $\mathbf{Q} (\br t)$ is defined by
\begin{equation}
\mathbf{Q} (\br t) = -i \langle \Psi ([v], t) | [\hat{\mathbf{j}} (\br), \hat{T} + \hat{W}] | \Psi ([v], t) \rangle \nonumber
\end{equation}
where $| \Psi ([v],t) \rangle$ is the time-dependent many-body state
obtained from the TDSE with potential $v$ and given initial state.
Eqs.(\ref{cont1}) and (\ref{cont2}) imply
\be
- \overrightarrow{\nabla} \cdot \left( n(\br t) \overrightarrow{\nabla} v(\br t) \right) = q([v] , \br t) - \partial_t^2 n(\br t).  \nonumber
\ee
where $q ( [v],\br t) = - \overrightarrow{\nabla} \cdot \mathbf{Q} (\br t)$ is regarded as a functional of $v$
through the state $| \Psi ([v],t) \rangle$. For a fixed density and initial state this is an implicit equation for the potential.
To solve this implicit equation we define an iterative sequence $v_k$ of potentials by the iterative solution of
\be
- \overrightarrow{\nabla} \cdot \left( n(\br t) \overrightarrow{\nabla} v_{k+1}(\br t) \right) = q([v_k] , \br t) - \partial_t^2 n(\br t) .
\label{fp}
\ee
In previous works \cite{GFPP,GFPP 1-D} we proved, for general initial states and interactions,
that under mild restrictions on the density the sequence $v_k$ converges in Banach norm sense to a potential $v$
which is both fixed-point of the equation and produces the prescribed density $n$.\\
Although the fixed point method itself is well-defined it is highly non-trivial
to develop a stable numerical algorithm. To do this we found it advantageous 
to make explicit use of also the
current (still being a functional of the density).  
This is most easily done for one-dimensional systems since the continuity Eq.(\ref{cont1}) can be integrated analytically.  
We find $j(xt)=j(at)+\tilde{j}(xt)$ where
\be
\tilde{j} (x t) = - \int_a^x dy \, \partial_t n(yt) ,
\label{cont3}
\ee
and where $a$ is an arbitrary point.
For this reason, and for simplicity of presentation, we restrict ourselves to the one-dimensional case in this Letter.
We first show how we can eliminate the quantity $q$ from our equations.
By integrating Eq.(\ref{fp}) and using Eq.(\ref{cont3}) we obtain
\be
- n(x t) \partial_x v_{k+1} (x t) = \partial_t \tilde{j}(x t) - Q ([v_k], xt) + c_1 (t) , \nonumber
\ee
where $c_1 (t)$ is an integration constant.
From Eq.(\ref{cont2}) for a system with potential $v_k$ we then find
\bea
- n(x t) \partial_x v_{k+1} (x t) &=& \partial_t \left[ \tilde{j}(x t) - j ([v_k],xt) \right]  \nonumber \\
&-&  n ( [v_k],xt ) \partial_x v_k (xt) + c_1 (t) .
\label{num1}
\eea
To obtain an equation that is only dependent on densities we can use Eq.(\ref{cont1}) and Eq.(\ref{cont3}) to find
\bea
- n(x t) \partial_x v_{k+1} (x t) &=&  \int_a^x dy  \, \partial_t^2 \left[ n([v_k],yt)  - n(yt)\right]  \nonumber \\
 &-&  n ( [v_k],xt ) \partial_x v_k (xt) + c_2 (t),
\label{num2}
\eea
where $c_2 (t)$ is a new constant. 
While mathematically equivalent Eqs.(\ref{num1}) and Eq.(\ref{num2}) are not numerically equivalent
as their discretizations on a space-time grid generally differ. 
In practice, it is therefore advantageous to use 
\bea
-n(xt) \partial_x v_{k+1} (xt) + n([v_k],xt) \partial_x v_{k} (xt) =  \nonumber  \\
 (1-\mu) \int_a^x dy  \, \partial_t^2 \left[ n([v_k],yt) - n(yt)  \right] \nonumber \\
+ \mu \, \partial_t \left[ \tilde{j}(x t) - j ([v_k],xt) \right] + c(t)  
\label{convex}
\eea
as follows immediately by multiplying Eq.(\ref{num1}) by $\mu$ and Eq.(\ref{num2}) by $1-\mu$ and
adding the results. Here $\mu$ is a parameter at our disposal and $c(t)$ is a new constant. 
This equation defines an iterative procedure to determine $v_{k+1}$ from $v_k$.
The constant $c(t)$ in this equation is uniquely determined by the spatial boundary conditions on $v_{k+1}$
(and hence depends on $k$).
When $v_k \to v$ then $n [v_k] \to n[v]$ and $j[v_k] \to j[v]$ and Eq.(\ref{convex})
implies
that $c (t) \to  \mu \partial_t j(at)$. Since we also obtain $|Ê\Psi ([v],t) \rangle$ after convergence
we can calculate any observable, and in particular the current $j(xt)$.
This is an explicit realization of the Runge-Gross result \cite{Runge-Gross} that any observable
is a functional of the density and the initial state. \\
{\em Numerical procedure.}
The iterative method based on Eq. (\ref{convex}) should be implemented stepwise in time 
for high efficiency. 
We use a midpoint based time-stepping method
which uses the midpoint potentials $\bar{v} (x t_n) = v (x, \frac{1}{2} (t_{n-1} + t_n) )$
to propagate the wave function on a time-grid with time-points $t_n$.
For this we implemented the Split Operator and Lanczos method \cite{Leforestier1991}.
Let us now suppose that we have obtained $\bar{v} (x t_m)$,
and hence the $| \Psi ([v ] ,t_m) \rangle$ giving the required density $n(x t_m)$ for $m \leq n$. 
Then to determine $\bar{v} (x t_{n+1})$, and hence $| \Psi ([v ] ,t_{n+1}) \rangle$,
we define an iterative procedure
in which we guess an initial potential and loop over potentials $\bar{v}_k (x t_{n+1})$ until we converge to the desired $\bar{v} (x t_{n+1})$:
\begin{enumerate}
\item
Use $\bar{v}_k (xt_{n+1})$ to calculate $| \Psi ([v_k ], t_{n+1}) \rangle$ from $| \Psi ([v ] ,t_n) \rangle$
by time-stepping. 
\item
From $|\Psi ([v_k], t_{n+1}) \rangle$  calculate $n([v_k] , x t_{n+1})$ and $j( [v_k] , x t_{n+1})$. 
\item
Calculate $\bar{v}_{k+1}  (x t_{n+1})$ from
\bea
\lefteqn{ - \bar{n} (x t_{n+1} ) \partial_x \left[ \bar{v}_{k+1} (x t_{n+1}) - \bar{v}_{k} (x t_{n+1}) \right] \Delta t^2 } \nonumber \\
&=& A \int_a^x dy \left[ n ( [v_k], y t_{n+1}) - n (y t_{n+1}) \right] \nonumber \\
&+& B\Delta t\left[ \tilde{j} (x t_{n+1})- j( [v_k],x t_{n+1}) \right] + c,
\label{UpN}
\eea
where $\Delta t = t_{n+1}-t_{n}$.
\end{enumerate}
Eq.(\ref{UpN}) is obtained from Eq.(\ref{convex}) by a discretization w.r.t. time 
using only times $t_m$ with $m \le n+1$ for the derivatives. We further used the fact that we have already converged up to time $t_n$
and replaced $\bar{n} [v_k]$ by $\bar{n}$ on the left hand side of the equation since we found that this does not affect the convergence.
The constants $A$ and $B$ depend on the discretization scheme and the $\mu$ of Eq.(\ref{convex}),
which effectively leaves the choice of their values at our disposal.
The constant $c$ in Eq.(\ref{UpN}) depends on the boundary conditions and hence the geometry of the system.
Below we will present examples for a periodic system. 
In that case the constant is determined by the periodicity condition $v_k (a t) = v_k (b t)$ on the spatial interval $[a,b]$.
This yields the condition
\begin{align}		
\int_a^b dx \left[ \frac{A\int_a^x dy \, \left[ n ([v_k], y t_{n+1}) - n ( y t_{n+1}) \right]}{\bar{n}(xt_{n+1}) }\right. \nonumber
\\
 + \left.  \frac{B\Delta t\left[ \tilde{j}( x t_{n+1}) - j ( [v_k], x t_{n+1}) \right] + c}{\bar{n} (xt_{n+1})}\right]  = 0.
 \label{condition}
\end{align}
During time-propagation numerical errors can build up in the modulus and phase of the wave function. The density is 
determined by the modulus only while the current also depends on the phase. This implies, for example, that errors
in the phase can lead to inaccurate currents while still producing an almost correct density. This tends to happen
in the case that $B=0$ since in that case the procedure enforces the correct density
without constraints on the current as can be seen from Eq.(\ref{UpN}). The opposite happens in the case that $A=0$. 
By taking nonzero values for $A$ and $B$ we control the accuracy of both the density and the current and
hence the modulus and phase of the wave function.
This suffices to stabilize the algorithm in most cases.
Perfect stability is obtained by spatially smoothening the potentials $\bar{v}_{k+1}$ as they are obtained.
Without smoothening the best algorithm is obtained when $A$ dominates $B$ for nonzero $B$ while
these values are not so important with smoothening (we used $A=1$ and $B=0.5$).
We find that $5$ iterations generally suffice to converge and that the precision of the potential
is limited mainly by the time-stepping method for the wave-function (assuming a sufficient spatial resolution). 
By increasing the precision thereof almost arbitrary precision can be achieved
even when the density changes by orders of magnitude.

\begin{figure}
\includegraphics[width=0.4\textwidth]{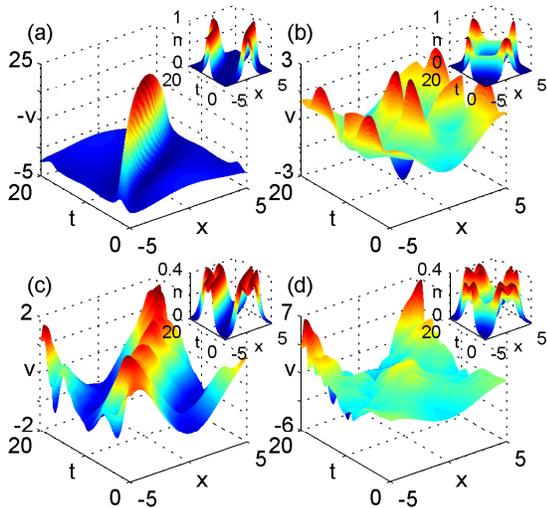}
\caption{(color online) The potentials that produce the prescribed densities $n_1$ and $n_2$ (insets).
Panel (a) $n_1, \lambda=0$,  (b) $n_2, \lambda=0$, (c) $n_1, \lambda=1$,  (d) $n_2, \lambda=1$.
Note that in (a) we plotted minus the potential for better visibility.}
\label{splittrans}
\end{figure}
{\em Translating and splitting a given density.}
To illustrate the algorithm we consider two electrons on a quantum ring of length $L=10$ over a time period of length $T=20$.
We start by calculating the singlet ground state $|\Psi_0 \rangle$ 
(which has a spatially symmetric wave function) 
of a (properly periodic) Hamiltonian with external potential $v_0$ and interaction $w$ given by
\begin{align}
 v_0 \left( x \right) =&  - \cos \left( {\frac{{2\pi x}}{L}} \right), \nonumber \\
 w\left( {{x_1},{x_2}} \right) =&  \lambda \cos \left( {\frac{{2\pi \left( {{x_1} - {x_2}} \right)}}{L}} \right) , \nonumber
\end{align}
where $\lambda$ is the interaction strength.
The ground state density is denoted by $n_0 (x)$.
We then construct the (spatially periodic) time-dependent densities $n_1$ and $n_2$ by:
\bea
n_1 (x t ) &=& n_0 (x-r(t)), \nonumber \\
n_2 (x t)  &=& \frac{1}{2} \left[ n_0 (x-r(t)) + n_0 (x+r(t)) \right], \nonumber \\
r\left( t \right) &=& \frac{L}{2}\left[ {1 - \cos \left( \frac{\pi t}{T} \right)} \right]. \nonumber
\eea
The density $n_1$ describes a situation where the initial density $n_0$ is rigidly translated around the ring exactly once
whereas the density $n_2$ describes a situation where the initial density $n_0$ is split in equal halves $\frac{1}{2} n_0$
that are rigidly translated in opposite directions to rejoin at times $\frac{1}{2} T$ and $T$.
We have used our algorithm to calculate the potentials that
produce these prescribed densities $n_1$ and $n_2$ via time-propagation of the initial state $| \Psi_0 \rangle$ by the TDSE.
This was done for the interaction strengths $\lambda=0$ and $\lambda=1$.
In Fig. \ref{splittrans} we present the corresponding potentials and densities (insets).
We see large differences in the potentials for the interacting case (panels (c) and (d))
as compared to the non-interacting case (panels (a) and (b)).
The convergence of our algorithm shows that the prescribed densities are indeed $v$-representable
and that the algorithm can be used for density changes of orders of magnitude.\\
\begin{figure}
\includegraphics[width=8.6cm]{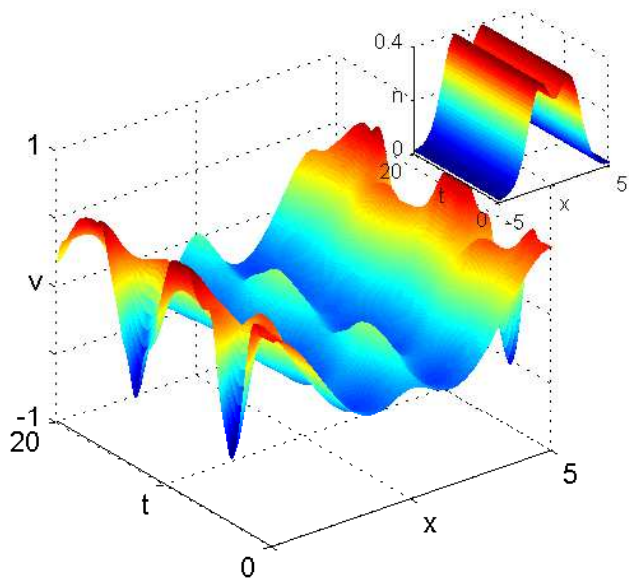}
\caption{(color online) The explicitly time-dependent KS potential that keeps the density (inset) static for the correlated initial state $| \Psi_0 \rangle$. We stress that the potential is not periodic in time.}
\label{groundstate}
\end{figure}
\begin{figure}
\includegraphics[width=8.6cm]{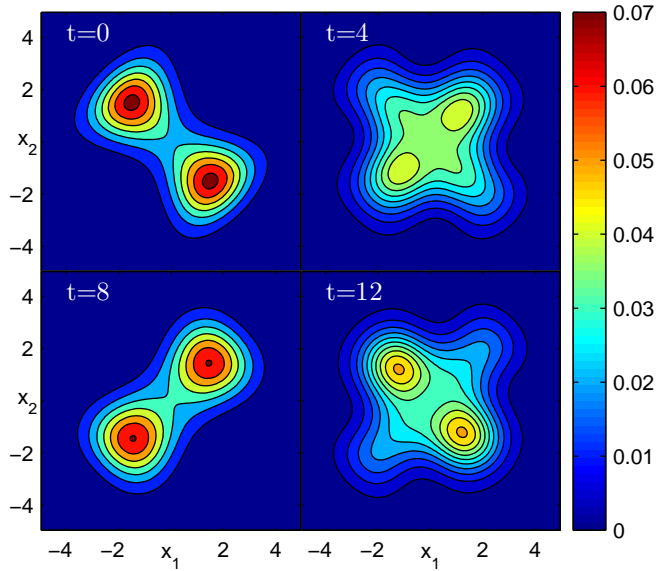}
\caption{(color online) 4 snapshots of $|\Psi_s (x_1,x_2,t)|^2$ at times where the KS potential in Fig.(\ref{groundstate}) is extreme. Note that the electrons are well-separated at time $t=0$ but are confined to the same region at time $t=8$.}
\label{snapshot}
\end{figure}
{\em Exact KS potential for a non-separable initial state.}
As a second example we construct an exact KS system, i.e.
a non-interacting system having the same time-dependent density as that of an
interacting reference system. 
For the KS system we also need to specify an initial state with the correct initial density $n_0$.
This state does not need to be the KS ground state (the ground state of a non-interacting system with density $n_0$)
as the Runge-Gross theorem \cite{Runge-Gross} allows for general initial states (see for further discussion \cite{ElliotMaitra}).
Here we take the KS initial state to be identical to the true correlated ground state
$| \Psi_0 \rangle$ of the interacting system. 
As the interacting reference system
we consider a system forever kept in the ground state $|\Psi_0 \rangle$ of the previous example for $\lambda=1$. 
The density is therefore stationary and equal to $n_0$.
Since $| \Psi_0 \rangle$ is not an eigenstate of a noninteracting system  
the KS state and potential will in general be time-dependent,
but in such a way that they still produce the static density $n_0$.
We denote the KS potential by $v_s$ and the KS Hamiltonian is thus given by
\be
\hat{H}_s (t) = -\frac{1}{2} \left(\partial_{1}^2 +\partial_{2}^2\right) + v_s (x_1 t) + v_s (x_2 t). \nonumber
\ee
We have determined the time-dependent potential $v_s$ with our algorithm and 
displayed it in Fig.\ref{groundstate}. The square $|\Psi_s (x_1,x_2,t)|^2$
of the corresponding KS wave function is
displayed in Fig.\ref{snapshot} at four times corresponding to extreme values of the KS potential. 
We see strong internal motions in the wave function
as it passes through states in which the electrons are well-separated $(t=0)$ and states where they are 
confined to the same region in space $(t=8)$,
although the corresponding density is completely static.		
The wave function at these times as well as the intermediate times $t=4$ and $t=12$ are in correspondence 
with the extreme values of the potential in Fig.\ref{groundstate}.
We also see the exact potential cannot be an adiabatic functional of the density, and hence must have memory,
as an adiabatic functional produces a static potential when we insert the exact density, in conflict with Fig.\ref{groundstate}.
This can be illustrated further by choosing as (the spatial part of) the initial KS state a separable state of the form
\be
\Psi_0 (x_1,x_2) = \phi (x_1) \phi (x_2) 
\label{init2}
\ee
with $\phi (x) = \sqrt{n_0(x)/2}$. In this case the KS-potential $v_s$ is static and given by
\be
\label{static}
v_s (x) = \frac{1}{2} \frac{\partial_x^2 \sqrt{n_0 (x)}}{\sqrt{ n_0 (x)}} 
\ee
up to an arbitrary constant. 
In this case the exact KS potential is static as would also have been predicted by any
adiabatic approximation.
This explicitly demonstrates the interplay between memory and initial states\cite{Maitra2002,MaitraBook}.\\
{\em Outlook.}
We presented a stable and fast algorithm to construct the external potential that, for a given initial state, 
produces a prescribed time-dependent density in an interacting many-body system.
The method will be valuable for further development of density functionals and local control theory.
Especially exciting is the possibility to use more advanced (multi-configurational) initial states in DFT
in combination with existing and new approximate functionals and to test them using our benchmarking algorithm.
This can open up new possibilities for the study of strongly correlated systems within a DFT framework.

{\em Acknowledgement.}
S.E.B.N. acknowledges support from the Lundbeck Foundation.
M.R. acknowledges support by the Erwin Schr\"odinger Fellowship J 3016-N16 of the FWF (Austrian Science Fonds). 
We further thank Prof. J. Olsen for valuable discussions.


\begin{thebibliography}{99}

\bibitem{DFT} R.M.\ Dreizler and E.K.U.\ Gross,
\textit{Density Functional Theory - An Approach to the Quantum Many-Body Problem} (Springer-Verlag, 1990).

\bibitem{Chayes1985} J.T.\ Chayes, L.\ Chayes, and M.B.\ Ruskai,
\href{http://www.springerlink.com/content/r44wr43860342183/}{J.  Stat. Phys. {\bf 38}, 497 (1985)}. 

\bibitem{Lammert2010} P.E.\ Lammert,
\href{http://pra.aps.org/abstract/PRA/v82/i1/e012109}{Phys. Rev. A {\bf 82}, 012109 (2010)}.

\bibitem{RvL1999} R.\ van Leeuwen,
\href{http://prl.aps.org/abstract/PRL/v82/i19/p3863_1}{Phys. Rev. Lett.  {\bf 82}, 3863 (1999)}.

\bibitem{Baer2008} R.\ Baer,
\href{http://jcp.aip.org/resource/1/jcpsa6/v128/i4/p044103_s1}{J. Chem. Phys. {\bf 128}, 044103 (2008)}

\bibitem{Li2008} Y.\ Li and C.A.\ Ullrich,
\href{http://dx.doi.org/10.1063/1.2955733}{J. Chem. Phys. {\bf 129}, 044105 (2008)}.

\bibitem{Verdozzi2008} C. \ Verdozzi,
\href{http://prl.aps.org/forward/PRL/v101/i16/e166401}{Phys. Rev. Lett. {\bf 101}, 166401 (2008)}

\bibitem{KurthS2011} S.\ Kurth and G.\ Stefanucci,
\href{http://www.sciencedirect.com/science/article/pii/S0301010411000346}{Chem. Phys. {\bf 391}, 164 (2011)}

\bibitem{CarstenBook} C. A.\ Ullrich,
\textit{Time-Dependent Density-Functional Theory} (Oxford University Press, 2012).

\bibitem{GFPP} M.\ Ruggenthaler and R.\ van Leeuwen,
\href{http://iopscience.iop.org/0295-5075/95/1/13001}{Euro Phys. Lett. {\bf 95}, 13001 (2011)}.

\bibitem{Farzanehpour2012} M.\ Farzanehpour and I.V.\ Tokatly,
\href{http://arxiv.org/abs/1206.6267}{arXiv:1206.6267v1 (2012)}.

\bibitem{Andrade2012} X.\ Andrade {\em et al.},
\href{http://iopscience.iop.org/0953-8984/24/23/233202}{J. Phys.: Condens. Matter {\bf 24}, 233202 (2012)}.

\bibitem{Burke2012} K.\ Burke,
\href{http://jcp.aip.org/resource/1/jcpsa6/v136/i15/p150901_s1}{ J. Chem. Phys. {\bf 136}, 150901 (2012)}.

\bibitem{Lein2005} M.\ Lein and S.\ K\"ummel,
\href{http://prl.aps.org/abstract/PRL/v94/i14/e143003}{ Phys. Rev. Lett. {\bf 94}, 143003 (2005)}.

\bibitem{Godby2012} J. D.\ Ramsden and R.W.\ Godby,
\href{http://prl.aps.org/abstract/PRL/v109/i3/e036402}{Phys. Rev. Lett. {\bf 109}, 036402 (2012)}.

\bibitem{GFPP 1-D} M.\ Ruggenthaler, K.J.H.\ Giesbertz, M.\ Penz, and R.\ van Leeuwen,
\href{http://link.aps.org/doi/10.1103/PhysRevA.85.052504}{Phys. Rev. A {\bf 85}, 052504 (2012)}.

\bibitem{OCT1} I.\ Serban, J.\ Werschnik, and E. K. U. Gross,
\href{http://link.aps.org/doi/10.1103/PhysRevA.71.053810}{Phys. Rev. A {\bf 71}, 053810 (2005)}.

\bibitem{OCT2} W.\ Zhu, J.\ Botina, and H.\ Rabitz,
\href{http://dx.doi.org/10.1063/1.475576}{J. Chem. Phys. {\bf 108}, 1953 (1998)};

\bibitem{LCT1} P.\ Gross, H.\ Singh, and H.\ Rabitz,
\href{http://link.aps.org/doi/10.1103/PhysRevA.47.4593}{Phys. Rev. A {\bf 47}, 4593 (1993)}.

\bibitem{LCT2} W.\ Zhu and H.\ Rabitz,
\href{http://dx.doi.org/10.1063/1.1582847}{J. Chem. Phys. {\bf 119}, 3619 (2003)}.

\bibitem{Mancini2005} S.\ Mancini, V.I.\ Manko, and H.M.\ Wiseman,
\href{http://iopscience.iop.org/1464-4266/7/10/E01/}{J. Opt. B: Quantum Semiclass. Opt. {\bf 7}, S177 (2005)}.

\bibitem{CHU02} S.\ Chu,
\href{http://www.nature.com/nature/journal/v416/n6877/full/416206a.html}{Nature {\bf 416}, 206 (2002)}

\bibitem{Runge-Gross} E.\ Runge and E.K.U.\ Gross,
\href{http://prl.aps.org/abstract/PRL/v52/i12/p997_1}{Phys. Rev. Lett. {\bf 52}, 997 (1984)}.

\bibitem{Leforestier1991} C.\ Leforestier {\em et al.},
\href{http://dx.doi.org/10.1016/0021-9991(91)90137-A}{J. Comp. Phys. {\bf 94}, 59 (1991)}.

\bibitem{ElliotMaitra} P. \ Elliott and N. \ Maitra,
\href{http://pra.aps.org/abstract/PRA/v85/i5/e052510}{Phys. Rev. A {\bf 85}, 052510 (2012)}

\bibitem{Maitra2002} N.T.\ Maitra, K.\ Burke, and C.\ Woodward,
\href{http://prl.aps.org/abstract/PRL/v89/i2/e023002}{Phys. Rev. Lett. {\bf 89}, 023002 (2002)}.

\bibitem{MaitraBook} N.T.\ Maitra,
\href{http://www.springerlink.com/content/f2176228603k971t/}
{Lect. Notes Phys. {\bf 837}, 167 (Springer, Berlin, Heidelberg, 2012)}.

\end{thebibliography}
\end{document}